\documentstyle[preprint,aps]{revtex}

\begin{document}

\title
{Analysis of proton-induced fragment production cross sections
by the Quantum Molecular Dynamics plus Statistical Decay Model}

\author{
Satoshi Chiba$^{(1,2)}$, 
Osamu Iwamoto$^{(2)}$,
Tokio Fukahori$^{(1,2)}$, 
Koji Niita$^{(1,3)}$,\\ 
Toshiki Maruyama$^{(1)}$,
Tomoyuki Maruyama$^{(1)}$
and 
Akira Iwamoto$^{(1)}$}

\address{
$^{1}$Advanced Science Research Center, 
Japan Atomic Energy Research Institute, \\
Tokai-mura, Naka-gun, Ibaraki-ken 319-11, Japan\\
$^{2}$Nuclear Data Center, 
Japan Atomic Energy Research Institute, \\
Tokai-mura, Naka-gun, Ibaraki-ken 319-11, Japan\\
$^{3}$Research Organization for Information Science 
and Technology \\
Tokai-mura, Naka-gun, Ibaraki-ken 319-11, Japan
}
\maketitle
\begin{abstract}

The production cross sections of various fragments from 
proton-induced reactions on 
$^{56}$Fe and $^{27}$Al have been analyzed by the Quantum 
Molecular Dynamics (QMD) plus Statistical Decay Model (SDM).
It was found that the mass and charge distributions calculated
with and without the statistical decay have very different
shapes.  These results also depend strongly 
on the impact parameter,
showing an importance of the dynamical treatment as realized
by the QMD approach.
The calculated results were compared 
with experimental
data in the energy region from 50 MeV to 5 GeV.     
The QMD+SDM calculation could reproduce the production
cross sections of the light clusters and 
intermediate-mass to 
heavy fragments in a good accuracy.   
The production cross section of  
$^{7}$Be was, however, underpredicted by 
approximately 2 orders of magnitude, showing 
the necessity of another reaction mechanism not taken into
account in the present model.

\end{abstract}



\section{Introduction}\label{introduction}

Recently we have developed a framework of the quantum
molecular dynamics (QMD) plus statistical decay model (SDM)
that takes account of the relativistic kinematics
and relativistic correction for interaction term, 
Lorentz boost of the initial and final states, 
realistic momentum distribution in the ground state, 
and comprehensive 
nucleon-nucleon (N-N) collision term\cite{niita95}.  
It was shown \cite{niita95}
that this framework could reproduce the 
measured double-differential
(p,xn), (p,xp') and (p,x $\pi$) reactions from 100 MeV to 3
GeV in a systematic way. In the subsequent 
papers \cite{chadwick95,chiba96}, we have given detailed
analyses of the pre-equilibrium (p,xp') and (p,xn) reactions
in terms of the QMD in the energy region of 100 to 200 MeV.  
It was demonstrated 
that the Fermi motion including
the high-momentum component, surface refraction, multi-step
effect and the multiple pre-equilibrium
emission were the key issues in understanding the 
angular distributions of neutrons and protons emitted 
during the pre-equilibrium process.  In these analyses  
a single set of parameters was selected, and no  
adjustment was attempted.  

The success obtained in the 
previous studies has shown the ability of the QMD+SDM
approach for the study of the nucleon-induced nuclear reactions.
However, the previous analyses have 
been concentrated on the inclusive particle spectra, and 
a fine selection of the final reaction products
was not performed.  
It is therefore the purpose of this 
work to carry out an analysis of proton-induced reactions 
for the production of specific final states, i.e., fragments,
with the 
same formulae and the same set of parameters as the
previous works\cite{niita95,chadwick95,chiba96}
to investigate further the validity of the
QMD+SDM approach.  
Such fragmentation phenomena themselves have been
a matter of long and intensive studies to 
extract the basic reaction mechanisms of
nucleon induced 
reactions
\cite{hufner85,barz86,botvina90,moretto93,gadioli,tanaka95}, 
e.g., multifragmentation,
liquid-gas phase transition, sideward peaking of
fragment angular distribution, 
properties of nuclei at
high temperature, etc. 
Reliable estimations of fragment production
are also important
in many application 
fields
\cite{nishida,sol,sihver93,vonach95,shigyo95,nagel,michel95}.

In this work,the 
fragment production cross 
sections from p + $^{56}$Fe and p + $^{27}$Al reactions
were calculated.
These targets were selected 
because
1) they are mono-isotopic or nearly mono-isotopic
in the natural elements, therefore 
the data are abundant, 
2) they are popular
elements among the structural materials with high 
importance
from the practical point of view, and 
3) effects of the fission and
multifragmentation 
become significant for heavier elements, where 
an analysis is too complicated.  
Moreover, the computation
time, proportional to the square of the mass
number, is kept manageable for these target 
materials. 
Special emapasis was placed for the proton 
energy of 1.5 GeV because this energy attracts 
special attention as a possible
candidate energy of application of a high-intensity
proton accelerator for researches on
transmutation of radioactive wastes and basic
sciences\cite{pec}.
The maximum energy was
chosen to be 5 GeV: above this energy many
nucleon-nucleon inelastic channles, which are not
considered in our model, are open.

In the next section, a brief explanation of the QMD plus
SDM approach is given.  The comparison of the calculation with
the experimental data and discussions on
the reaction mechanisms are given in section \ref{results}. 
Summary of this work is given in section \ref{summary}.

\section{Essence of the QMD plus SDM Model}
\label{brief}

The details of the QMD and SDM calculations are given in 
Ref. \cite{niita95}, so only 
the basic principles of them will be repeated here.
In the QMD calculation, each nucleon is expressed with a
Gaussian wave packet 
in both the coordinate and momentum spaces in the following way:

\begin{equation}
f_i({\bf r}, {\bf p}) = 8 \cdot \exp \left[ -\frac{({\bf r}-{\bf 
R}_i)^2}{4L} 
-\frac{2L(\bf{p}-\bf{P}_i)^2}{ \hbar ^2} \right]
\label{eq:phase-space}
\end{equation}
where ${\it L}$ is a parameter which represents the spacial spread of 
a wave packet, 
${\bf R}_{\it i}$  and ${\bf P}_i$ corresponding to
the centers of a wave packet in the coordinate and momentum spaces, 
respectively.  The total 
one-body phase-space distribution function is taken to be simply a sum
of these single-particle wave packets.  
Initially, we distribute the 
${\bf R}_{\it i}$  and ${\bf P}_i$ to produce a stable
target ground state with realistic density and momentum distributions.
The time evolution of 
${\bf R}_{\it i}$ and ${\bf P}_i$ 
is determined 
based on the Newtonian equation
and the N-N collision term, latter satisfying the Pauli
exclusion principle.  
A Skyrme-type interaction parameterized in Ref. \cite{niita95}
is used as the effective interaction.  In addition,
the symmetry and the Coulomb forces are included.

The QMD calculation is carried out up to a time scale 100[fm/c].
The position and momentum of each 
nucleon is then used to calculate the distribution of 
mass and atomic numbers, 
kinetic energy and direction of motion of the 
remaining fragments (which are referred to as "prefragments") 
as well as those of the emitted nucleons
and $\pi$-mesons.    
In determining the mass and atomic 
numbers of the prefragments,
the phase-space
minimum-distance-chain method\cite{maruyama92} is employed.

The prefragments thus identified are then
Lorentz boosted into their rest frames to evaluate 
their excitation energies.
When the prefragment is in the excited state,
the statistical decay via $n$, $p$, $d$, $t$, $^{3}He$ and $\alpha$ 
emissions is considered based on the Weisskopf-Ewing approximation, 
the emission probability of particle $x$ being given as

\begin{equation}
P_{x}=(2J_{x}+1)m_x \epsilon \sigma_x(\epsilon) \rho(E) d\epsilon
\label{eq:we}
\end{equation}
where $J{_x}$, $m{_x}$ and $\epsilon$ are the spin, mass and kinetic
energy of the emitted particle, while $\sigma_{x}(\epsilon)$ and
$\rho(E)$ denote the inverse cross section and the level density
of the residual nucleus at the excitation energy $E$, respectively.
The level density has been assumed to be proportional to
$\exp(2\sqrt{aE})$ with $a=A/8$ MeV$^{-1}$.  The inverse 
cross section is taken to be of the form
$\sigma_{x}(\epsilon)=(1-U_{x}/\epsilon)\pi R^2$ if 
$\epsilon > U_{x}$ and $\sigma_x(\epsilon)=0$ 
otherwise, where $R$ denotes the 
absorption radius and $U_{x}$ is the empirical Coulomb
barrier for particle $x$\cite{charity88}.

The separation of the QMD and SDM calculations
as performed in this approach
can give the individual  
production cross sections of various residues before
and after the statistical decay, the former being called as
the prefragment, the latter as the fragment.  
This possibility gives information 
on the relation of the dynamical and statistical
processes as a function of projectile energy and impact
parameter.
Calculations with event
number of 50,000 for each combination
of target and proton energy have been performed 
to get enough statistical 
accuracy. 

\section{Results and Discussion}
\label{results}

Calculated cross sections for the production of various 
prefragments and
fragments are shown for 1.5-GeV p + $^{56}$Fe
reaction in Fig. 1.
Here the upper figure
shows the production cross sections calculated only by the 
QMD, while the lower one is obtained including the 
statistical decay.  In this and the subsequent 
figures, the cross sections before and after the 
statistical decay process is
denoted as the "QMD" and "QMD+SDM",
respectively.
The small squares indicate the positions of stable nuclei, while
the lines at $N,Z$=20 and 28 show the locations of major magic
numbers.
The upper part of Fig. 1 shows that the prefragments are 
produced primarily in mass $A$ = 1 $\sim$ 10 
and $A_{T}$/2 $\le$ $A$ $\le$ $A_{T}$
regions, with $A_{T}$ as the target mass.  
In the cascade model approach\cite{bertini,nakahara},
the light mass fragments ($A$ $\le$ 10) 
are produced only as a result of the 
statistical decay or as a spallation residue, whereas 
the QMD describes the dynamical emission of such
light mass fragments 
as shown in Fig. 1.
This is a clear advantage of the QMD approach.
The intermediate-mass-fragment (IMF) corresponding 
to mass 10 to $A_{T}$/2 are not represented by the QMD 
result alone.

In the target region, the 
prefragments are distributed in a broad area of the 
N-Z plane.  This distribution,
then, is changed via the statistical decay to the 
bottom figure which is more localized 
along the stability line.  The IMF region 
is filled by the statistical decay of heavier prefragments,
and the final isotopic distribution becomes significantly flattened
in the direction of stability line
as a result of the statistical decay.  At the same time, yield
of the light mass fragment becomes much larger, especially for 
$\alpha$ particle.  Figure 1 demonstrates clearly the importance 
of both the dynamical and statistical processes which are
included in the QMD+SDM approach.

Figure 2 shows the total mass (top figure) and 
charge (bottom figure) distributions from 
the QMD and QMD+SDM calculations for the same projectile/target
combination.  
This figure again confirms that the QMD calculation predicts
dynamical production 
of the light ($A$ $\le$ $\sim$ 10) and target mass/charge
regions.  The statistical decay then makes the distributions much
smoother and flatter.  The IMF is produced primarily by the 
statistical decay.  

In Fig. 3, the mass distributions obtained for 1.5-GeV
p + $^{56}$Fe reaction with different impact parameter
events are shown.  
In this figure the impact parameter (denoted as $b$)
varies from 0 to 1 fm in the left-top figure, then
the impact parameter range is increased in
steps of 1 fm toward the right-bottom one. 
The total mass-yield distributions given in the upper part
of Fig. 2 were obtained by integrating the contributions
from the whole impact parameter range given in Fig. 3.  
Figure 3 indicates that the mass distribution changes 
significantly as a function of the impact parameter.  
For the events having smaller impact parameters, the
QMD distribution has a broad maximum centered around a 
mass of
47.  Then this distribution becomes much broadened by the 
statistical decay, shifting the peak to around 
mass $\sim A_{T}$/2.  As the impact parameter increases, 
the distribution is shifted toward higher masses;
at the peripheral events (b $\ge$ 4 fm) the QMD distribution
has a sharp peak at the target mass.  In this impact
parameter region, 
the incident proton interacts with a nucleon by grazing 
the target
near the surface. Therefore the chance that one of these 2
nucleons is
emitted from the target without further collisions 
is very high
due to a smaller nucleon density than in
the central region.  For such a reaction, which may be
referred to as "1-step" reaction, the cross section 
is determined by 
the magnitude of the basic nucleon-nucleon cross section.
Furthermore, Fig. 3 shows that the rather smooth and 
monotonic distributions 
seen in Figs. 1 and 2 are the results of superpositions of
contributions from different impact parameters which 
have quite
different shapes. Therefore, the dynamical approach
as realized by QMD is essential in predicting such
fragment distributions.

In Fig. 4 shown are the production cross sections
of various fragments for 1.5-GeV p + $^{56}$Fe reaction.
The data were measured by Michel et 
al\cite{michel95} at 1.6 GeV for $^{nat}$Fe.  
It is clearly concluded that the QMD+SDM
approach reproduces the fragment production cross sections
in the whole mass region well, 
including the light clusters 
such as $\alpha$ and IMF ($A \sim$ 20 to 30).  
However, the production cross section of  
$^{7}$Be is  
underestimated by approximately 2 orders of magnitude.
The multifragmentation, which is
not included in the calculation, may be
the primary source to resolve this very interesting 
issue\cite{michel95}.

The calculated  
isotope production cross sections 
from proton induced
reaction on $^{56}$Fe and $^{27}$Al 
are compared with experimental data in Figs. 5 and 6 for various 
residual nuclei as a function of incident energy 
in the energy region of 50 MeV to 5 GeV.  
We did not distinguish the experimental data for $^{56}$Fe and
$^{nat}$Fe in order to enhance the experimental database. 
In these figures, the results of the
QMD calculation
is given by the broken curves, the QMD+SDM
result 
by the full curves, the experimental data by the open 
circles with error bars.  The error bars in the QMD and QMD+SDM
calculations indicate the statistical uncertainty.
The experimental data have been retrieved from the 
CHESTOR (Charged Particle Experimental Data Storage and 
Retrieval System) database at Nuclear Data Center of JAERI
\cite{fukahori} supplemented by available 
literatures including the recent data reported by
Michel et al\cite{michel95}. 

The $^{56}$Fe(p,x) cross sections shown in Fig. 5 again reveals the
general accuracy of the QMD+SDM approach for the a-priori
estimation of production cross sections of target-like fragment
for a very wide incident energy range;
the QMD+SDM calculation reproduces the 
experimental data within a factor of 2.  
In most cases, shapes of the prefragment production cross sections
(the broken curves) are quite different from those of the final cross 
sections (solid curves) except for
such a simple reaction as the (p,n) case where the 
prefragment production
cross section 
is always larger than the final cross section by a 
constant ratio.  The difference between the "QMD" and "QMD+SDM"
calculations shows that the production of these
fragments is a result of subtle balance between the dynamical 
formation and
the statistical decay processes.  
The dynamical process acts as a "source"
of "hot" prefragments at the target mass region, 
followed by the production of nuclides with 
smaller masses due to particle evaporation.
It is intersting to note that the calculated 
$^{56}$Fe(p,n)$^{56}$Co
reaction cross section increases at energies above 
several hundred MeV.
The same energy
dependence was obtained for 
$^{56}$Fe(p,2n)$^{55}$Co and 
$^{59}$Co(p,n)$^{59}$Ni reactions.
It is natural to assume that such reactions
leading to the formation of prefragments which are 
very close to the target
take place mostly at the peripheral region (as 
Fig. 3 clearly demonstrates). 
For these events the number of collisions 
in the compound system may be 
only 1 or 2.
Then, the cross section for such events will be roughly 
proportional to the basic nucleon-nucleon cross sections
adopted in the calculation.
Indeed, the p-n
cross section used in our calculation increases
at the energy above 400 MeV 
due to contributions of the inelastic channels
(see Fig. 1 of Ref. \cite{niita95}).
On the contrary, the recent data measured by 
Michel et al.\cite{michel95} 
shows a steep drop above 1 GeV. 
However,
the $^{nat}$Fe(p,xn)$^{57}$Co and 
$^{nat}$Fe(p,xn)$^{58}$Co reactions reported by them
have a clear increase above several hundred MeV, although they 
attribute this energy dependence to the influence
of secondary reactions.  We hope that further 
experimental investigation 
of such reactions would be carried out to obtain 
information on this subject because it could be a direct 
measure of the N-N inelastic collision in nuclei.

Figure 6 exhibits a result of the similar analysis
for $^{27}$Al target.  
The cross sections 
for production of target-like fragments are reproduced
very well by the QMD+SDM calculation.  
On the contrary,
the production cross section of 
$^{7}$Be is noticeably
underestimated, even the threshold energy not being 
reproduced correctly.  These results are consistent with
the case for p+Fe reaction as shown in Fig. 4.  
Other reaction mechanisms which are not taken into
the present model, including the 
multifragmenttaion,
might be the origin of production of such 
"heavy" clusters as $^{7}$Be.

\section{Summary}\label{summary}

We have calculated the production cross sections
of various residues for 1.5-GeV proton-induced 
reactions on $^{56}$Fe in terms 
of the Quantum Moleculear 
Dynamics (QMD) and the Statistical Decay Model 
(SDM).  It was found that the distribution of 
the fragments calculated by QMD alone and QMD+SDM
differs considerably.  The distribution after the
QMD calculation alone has a broad maximum close
to the target mass and a maximum at 
the very light mass region.  This distribution 
is then smoothed out by the statistical decay
along the stability line, filling the gap at the 
intermediate-mass-fragment (IMF) region.

The QMD calculation
predicts the dynamical production of light 
fragments which is not possible
with the ordinary cascade model approach.  Furthermore,
it was found that the distribution of the fragments
depends strongly on the impact parameter, showing
the importance of the dynamical treatment
as realized by QMD. 

The calculated results for proton-induced 
fragment production cross sections on $^{56}$Fe and 
$^{27}$Al have been compared with experimental data
in the energy range of 50 MeV to 5 GeV.  A 
satisfactory overall agreement of the QMD+SDM calculation with
the measured data is obtained, which 
confirmed the basic validity of
the model and underlying parameters adopted 
in the present approach.  The production cross section of 
$^{7}$Be, however,
showed that other production mechanisms  
which are not included in this model 
may be needed to improve the 
agreement between the theory and experimental data.
This problem, together with the high energy behavior
of the $^{56}$Fe(p,n)$^{56}$Co reaction, should be 
investigated further for a better understanding of the 
phenomena.

\acknowledgements

Th authors wish to thank  
Drs. Hiroshi Takada and Peter P. Siegler of JAERI,
and Dr. Shiori Furihata of Mitsubishi Research Institute
for valuable discussions and comments.


\begin{figure}
\caption{Distribution of various 
fragments in N-Z plane
from the QMD alone
(top)
and QMD+SDM(bottom) calculations 
for 1.5-GeV proton + $^{56}$Fe reaction.  The small
squares indicate positions of stable isotopes, while
vertical and horizontal lines at N,Z=20 and 28 denote
the magic numbers.
}
\end{figure}

\begin{figure}
\caption{Calculated mass (top) and charge (bottom) 
distributions of fragments for 1.5-GeV p +
$^{56}$Fe reaction.  The broken histograms
show the results for QMD calculation only, while
the solid ones corresponding to QMD+SDM results.}
\end{figure}

\begin{figure}
\caption{Calculated mass distributions for 
1.5-GeV p + $^{56}$Fe reaction with different
impact parameters.}
\end{figure}

\begin{figure}
\caption{Production cross sections of various fragments for
1.5-GeV proton + $^{56}$Fe reaction.  The full circles
connected by a solid line denote the result of QMD+SDM
calculation, while the open circles connected by
a broken line were obtained experimentally by Michel 
et al. measured at 1.6 GeV for $^{nat}$Fe}
\end{figure}

\begin{figure}
\caption{Calculated and measured fragment production
cross sections for p + $^{56}$Fe reaction as a 
function of incident energy.}
\end{figure}

\begin{figure}
\caption{Same as Fig. 5 but for p + $^{27}$Al 
reaction.~~~~~~~~~~~~~~~~~~~~~~~~~~~~~~~~~~~~~~~~~~~~~~~~~~~~~~~~}
\end{figure}

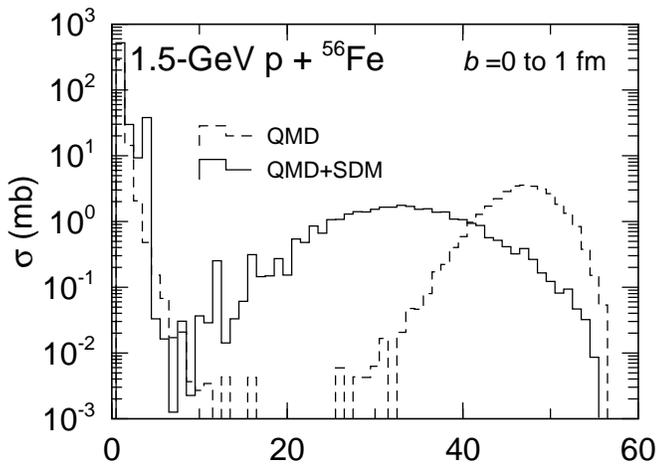
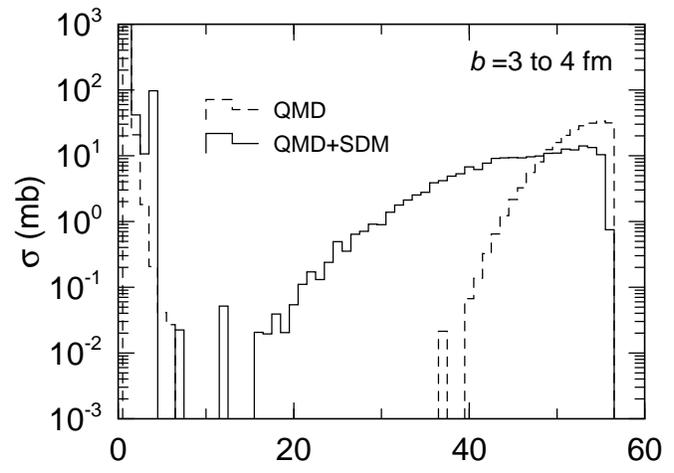
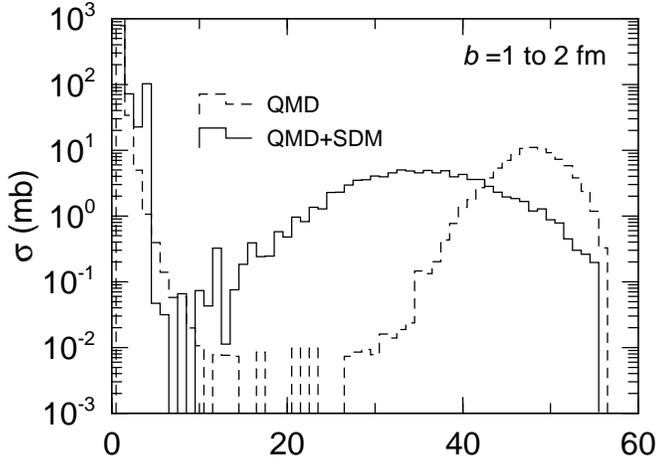
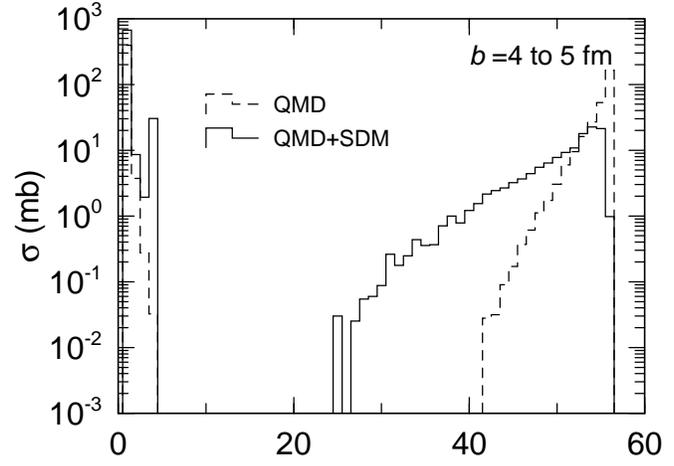
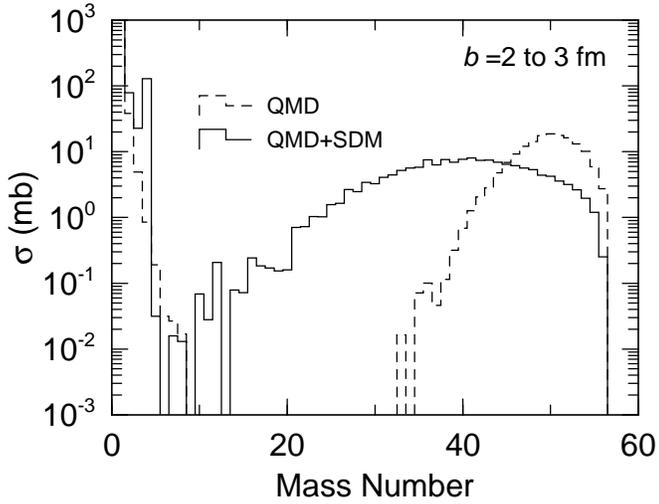
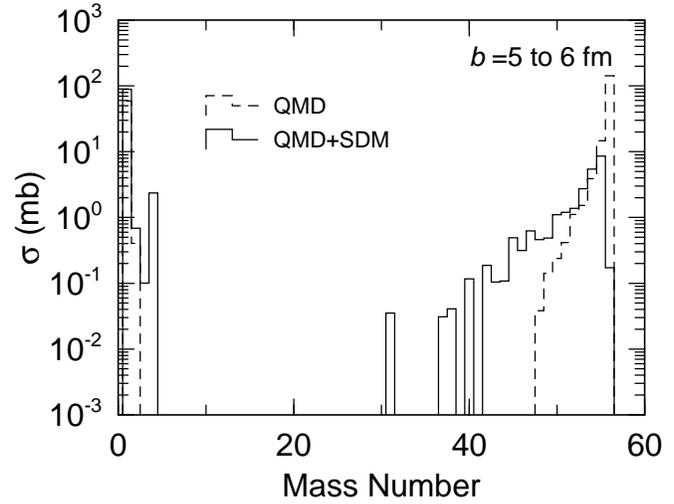
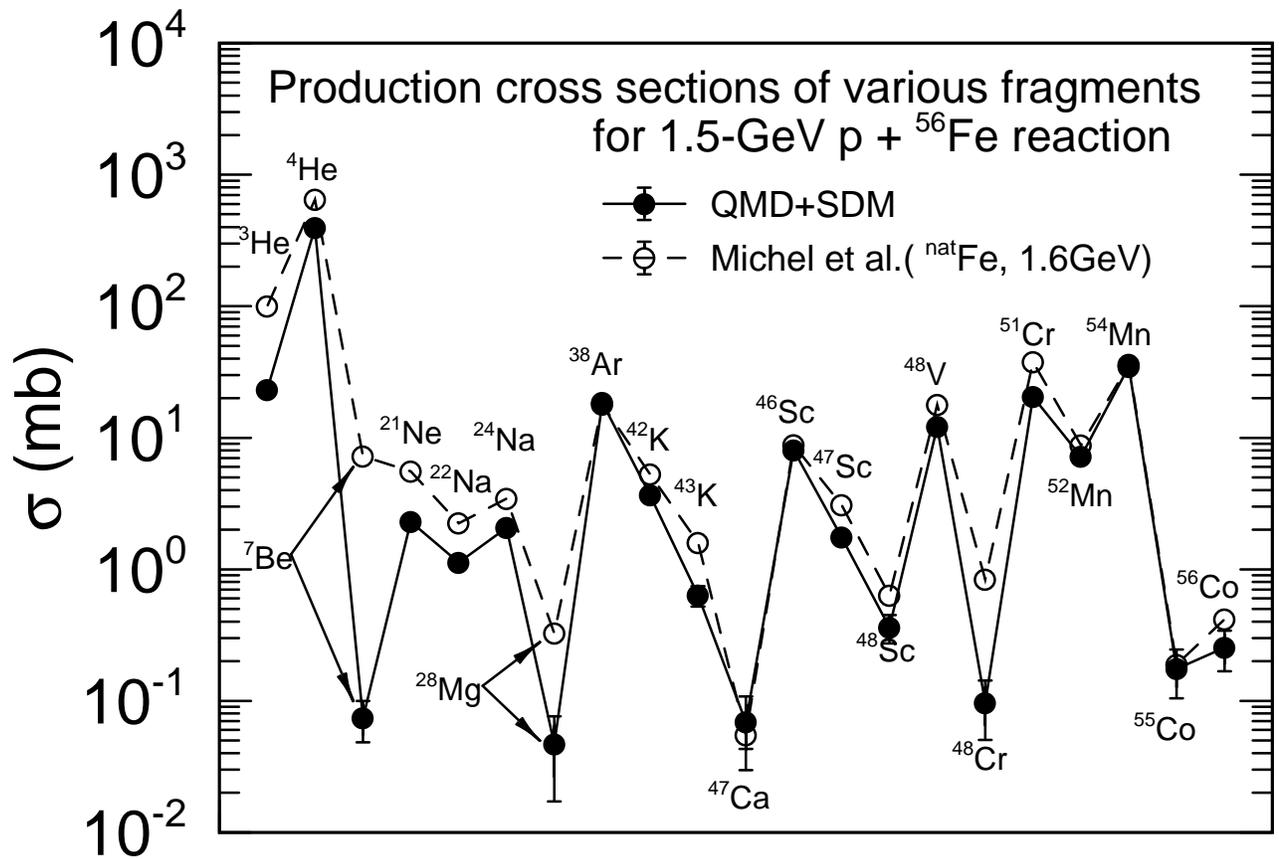
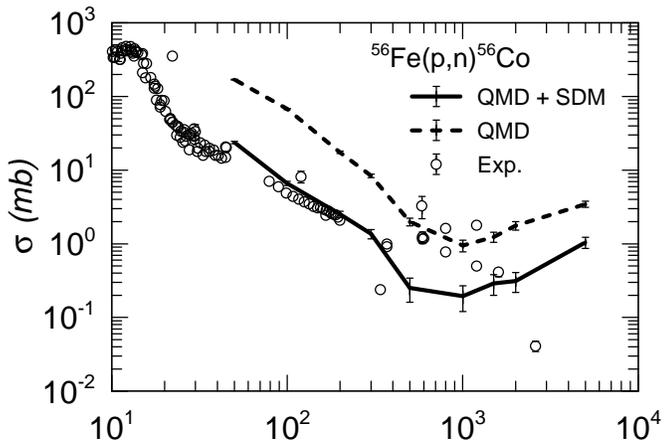
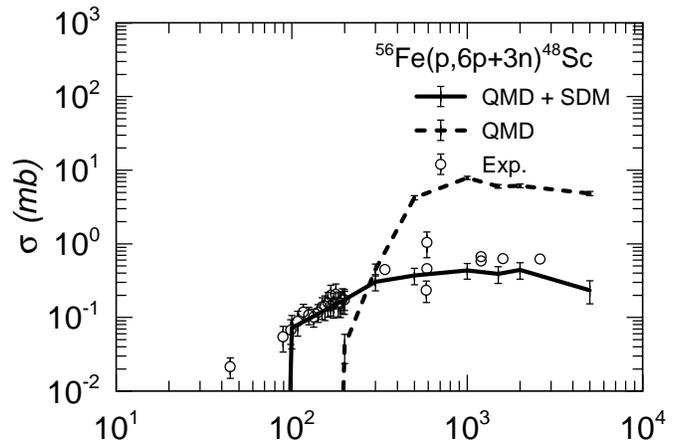
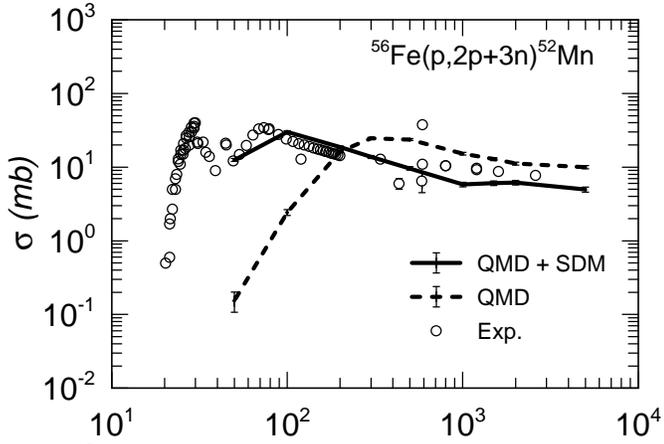
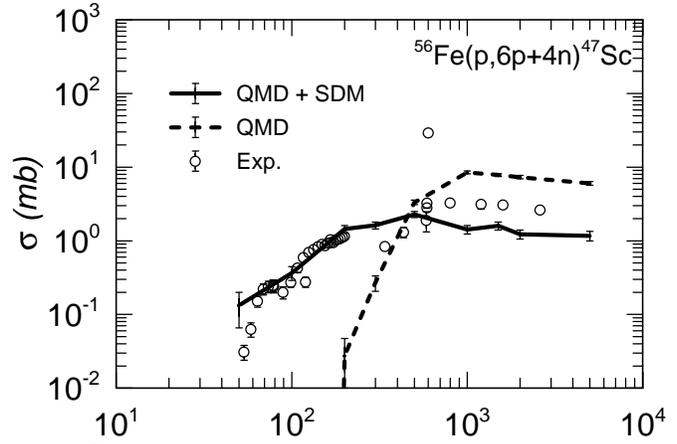
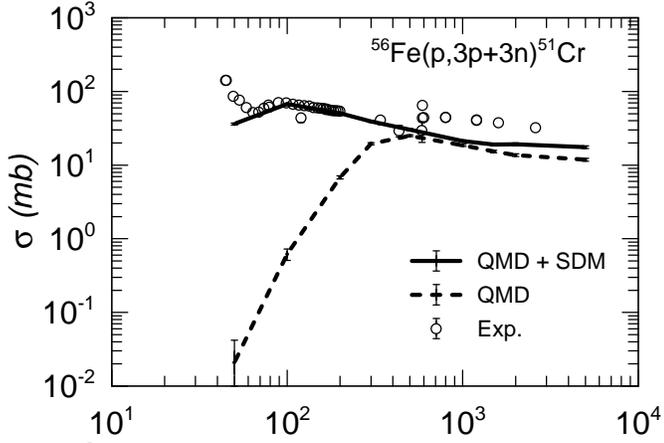
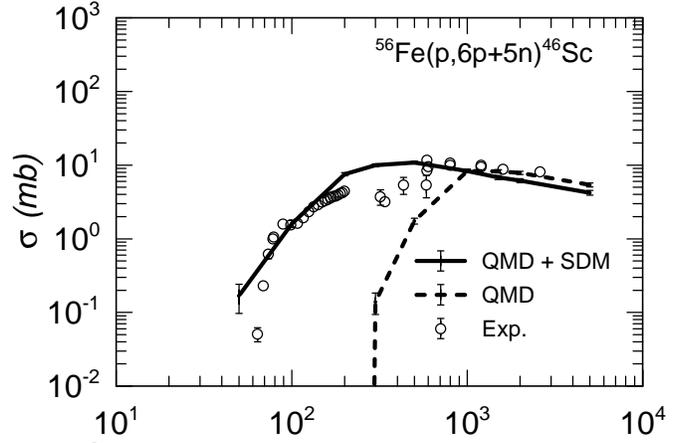
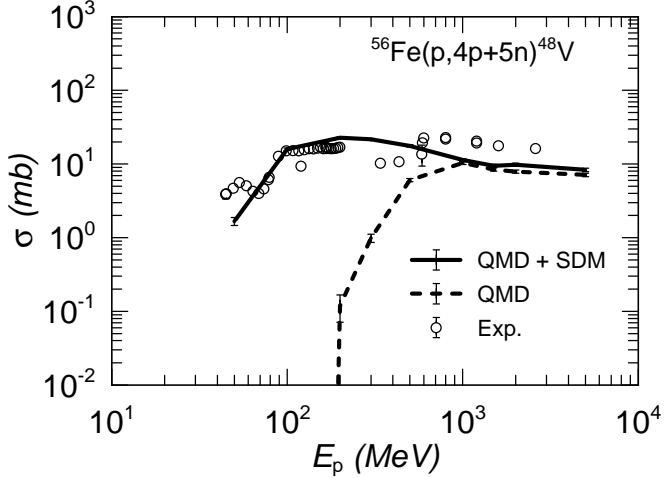
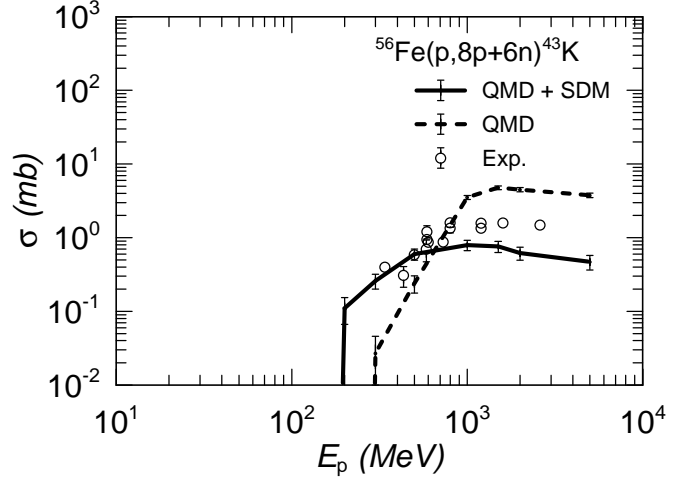
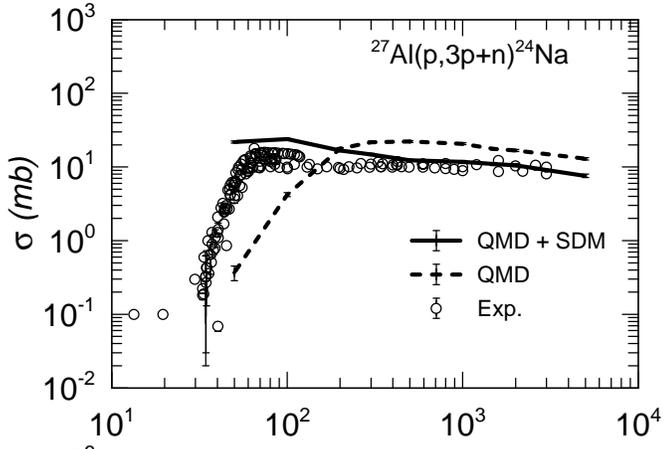
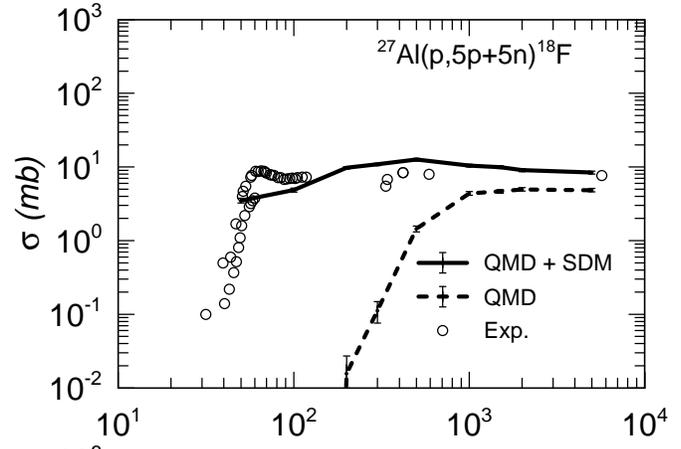
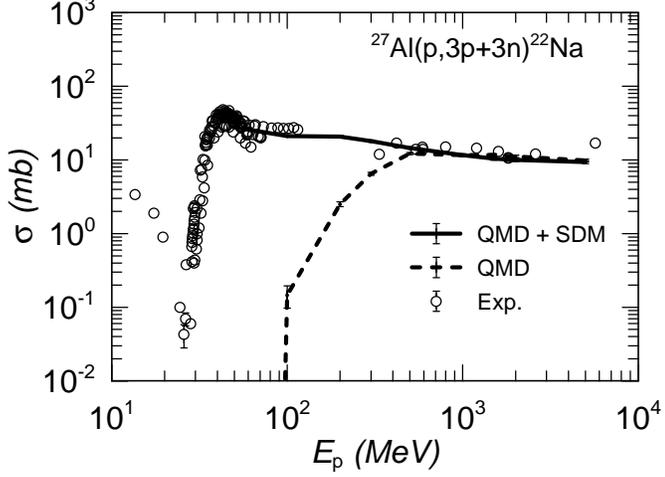
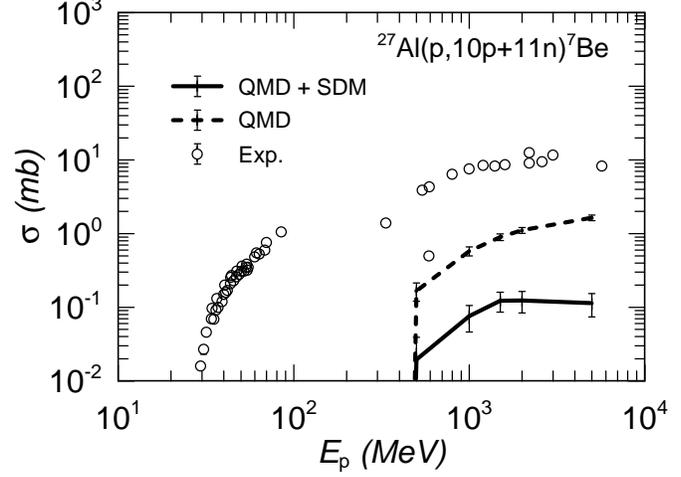
\end{document}